\documentclass[pra,aps,showpacs,twocolumn, a4paper,tightenlines,balancelastpage]{revtex4}
\usepackage{mathrsfs}
\usepackage{amsfonts}
\usepackage{amsmath,amsfonts,amssymb,mathrsfs,bm}
\usepackage{verbatim,psfrag,times,CJK}
\usepackage{graphicx,graphics,color,epsfig}
\usepackage{float}
\usepackage{flafter}
\usepackage{subeqnarray}
\usepackage{cases}
\usepackage{amsmath, upgreek, amssymb, graphicx, subfigure, paralist, dsfont}
\usepackage[colorlinks, linkcolor=blue, citecolor=blue, urlcolor=blue, breaklinks=true]{hyperref}

\definecolor{RED}{rgb}{1,0,0}
\newcommand{\ket}[1]{\left\vert{#1}\right\rangle}

\newcommand{\av}[1]{\langle{#1}\rangle}

\begin{document}

\title{Engineering a squeezed phonon reservoir with a bichromatic driving of a quantum dot}
\author{Bo Gao}
\author{Gao-xiang Li}
\email{gaox@phy.ccnu.edu.cn}
\affiliation{Department of Physics, Huazhong Normal University, Wuhan 430079, P. R. China}
\author{Zbigniew Ficek}
\affiliation{The National Center for Applied Physics, KACST, P.O. Box 6086, Riyadh 11442, Saudi Arabia}

\begin{abstract}
We demonstrate how an acoustic phonon bath when coupled to a quantum dot with the help of a bichromatic laser field may effectively form a quantum squeezed reservoir.
This approach allows one to achieve an arbitrary degree of squeezing of the effective reservoir and it incorporates the properties of the reservoir into two parameters, which can be controlled by varying the ratio of the Rabi frequencies of the bichromatic field. It is found that for unequal Rabi frequencies, the effective reservoir may appear as a quantum squeezed field of ordinary or inverted harmonic oscillators. When the Rabi frequencies are equal the effective reservoir appears as a perfectly squeezed field in which the decay of one of the polarization quadratures of the quantum dot dipole moment is inhibited. The decay of the quantum dot to a stationary state which depends on the initial coherence is predicted. This unusual result is shown to be a consequence of a quantum-nondemolition type coupling of the quantum dot to the engineered squeezed reservoir. The effect of the initial coherence on the steady-state dressed-state population distribution and the fluorescence spectrum is discussed in details. The complete polarization of the dressed state population and asymmetric spectra composed of only a single Rabi sideband peak are obtained under strictly resonant excitation.
\end{abstract}
                                                                                                                                                                                                                                                                                  

\date{\today}

\maketitle

\section{introduction}\label{sec1}

The study of the effect of phonons on the dynamics and coherent excitation of a quantum dot has been the subject of considerable interest in recent years. A number of different situations have been investigated~\cite{rm08,mb12,rgg10,mt13,fw03,ah04,mn10,vc11,md11,mp12,mn13,wu12,zh13,zh14,yc11}. These include experimental studies of the effect of phonons on the Rabi oscillations, Autler-Townes splitting, and the Mollow triplet of the fluorescence field emitted by a driven quantum dot~\cite{ka06,vc07,fm09,vz09,rg10,ua11,rh11,rh12,hh15}. In particular, it has been observed that the linewidths of the Rabi sidebands of the Mollow triplet increase linearly with temperature and with the square of the driving field strength~\cite{uw13,wh13}. These properties of the spectrum have been explained as arising from the coupling of the exciton transition of the quantum dot to longitudinal acoustic phonons~\cite{kamp}. Moreover, an interesting phenomenon of population inversions between the excitonic states of a quantum dot located inside an optical cavity and interacting with a phonon bath has been demonstrated both theoretically and experimentally~\cite{hc13,macovei,huang,simon,wu}. The investigation of this interaction in a quantum dot-cavity system has led to the prediction of single photon sources and the realization of single-photon devices~\cite{singlephoton1,singlephoton2,sci,ding}. The influence of a phonon bath on the photon blockade effect in a driven dot-cavity system and the emission of correlated and entangled photons has also been treated~\cite{dressed1,gao,na08}.

It is well known that the decay of a quantum system can be controlled and significantly modified by coupling the system to a squeezed vacuum field, which is characterized by the noise in one of the field quadrature components reduced below the usual vacuum level~\cite{ga86,cl87,pa90,df99,df04}. Particularly interesting effects of the squeezed vacuum field on the atomic radiative processes are the inhibition of the atomic decay and its dependence on the squeezing phase.
However, there are many practical problems with the application of the squeezed field produced by an external sources~\cite{gp95,gp97,wk86,wx87,mw13,te16}. The main obstacle is due to the requirement
that nearly all the modes to which the quantum system is coupled must be squeezed. In addition, there is a general lack of squeezed light sources coinciding with convenient atomic transitions~\cite{tg98}. It has been proposed that these difficulties could be circumvent by engineering a squeezed-reservoir-type interaction of a quantum system rather than coupling the system to a squeezed field produced by an external source~\cite{pc96,cp03}. For example, L\"{u}tkenhaus {\it et al.}~\cite{lc98} have studied the dynamics of a four-level system driven by two laser fields and have shown that the system may effectively behave as a two-level system coupled to a squeezed reservoir.

In this paper we propose a method to construct a squeezed-vacuum type multimode reservoir from a phonon bath based on a suitable engineering of the coupling of the phonon bath to a quantum dot. The quantum dot is modeled as a two-level system and experiences fluctuations and decay of its excitation that are due to the dissipative interaction with the phonon bath. We derive the master equation for the reduced density operator of the quantum dot and show that the phonon bath combined together with a bichromatic laser field tuned close to the dot's transition frequency can result in a squeezed reservoir type interaction of the phonon bath with the quantum dot. We find that the squeezing properties of the effective reservoir and then the quantum dot's relaxation dynamics can be controlled through variation of certain tunable system parameters, e.g., the Rabi frequencies of the bichromatic field. By varying the ratio between the Rabi frequencies of the bichromatic field the effective squeezed reservoir displays interesting differences in its properties. In particular, the reservoir may appear as a quantum squeezed field of ordinary or inverted harmonic oscillators, or can behave as a perfectly squeezed field. When in addition to the interaction with the squeezed reservoir, the quantum dot is driven by a resonant laser field we find that the steady-state dressed state population distribution and in the properties of the fluorescence field can be governed by the initial coherence between the ground and excited states of the quantum dots. The fluorescence spectrum can be asymmetric and its structure varied with the initial coherence. We show that the asymmetries are manifestation of the complete polarization of the dressed state populations, and thus the spectrum offers a method of observing the polarization.

The paper is organized as follows. In Sec.~\ref{sec2}, we describe the model and derive the master equation for the reduced density operator of a quantum dot interacting with a low frequency phonon bath and driven by a bichromatic laser field. In Sec.~\ref{sec3}, we examine the conditions for quantum features of the engineered reservoir and their dependence on the number of phonons. We distinguish between different forms of the squeezed phonon reservoir which can be engineered, including a perfectly squeezed reservoir and a squeezed reservoir of inverted harmonic oscillators.
In Sec.~\ref{sec4}, we concentrate on the dynamics of the quantum dot which in addition to the interaction with the engineered reservoir is driven by a resonant laser field.
We are particularly interested in the stationary state and its dependence on the form of the engineered squeezed reservoir. Section~\ref{sec5} is devoted to the discussion of the fluorescence spectrum. The dependence of the stationary spectrum on the initial coherence is exhibited and explained in terms of the dressed states of the driven quantum dot. The results are summarized in Sec.~\ref{sec6}.

\section{Description of the system}\label{sec2}

\begin{figure}[h]
\centering\includegraphics[width=4cm,keepaspectratio,clip]{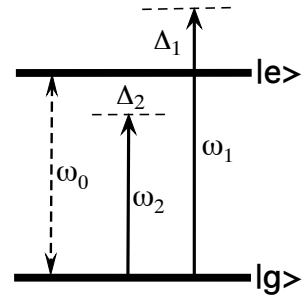}
\caption{Two-level system driven by a bichromatic laser field of frequencies $\omega_{1}$ and $\omega_{2}$ tuned close to the atomic transition frequency~$\omega_{0}$ at detunings $\Delta_{1}$ and $\Delta_{2}$, respectively. }
\label{fig1}
\end{figure}

We consider a single quantum dot (QD) coupled to a low frequency phonon bath and driven by a bichromatic laser field.
The quantum dot is modeled as a two-level system with the upper state $\ket e$, the ground state $\ket g$, transition frequency $\omega_{0}$, and transition dipole moment $\vec{\mu}$. The driving field is characterized by two frequencies $\omega_1$ and $\omega_2$ and the amplitudes $\vec{\mathcal{E}}_{1}$ and $\vec{\mathcal{E}}_{2}$, respectively. The components of the laser field are tuned near the atomic resonance, at detunings $\Delta_{1}=\omega_{1}-\omega_{0}$ and $\Delta_{2}=\omega_{0}-\omega_{2}$, as illustrated in Fig.~\ref{fig1}. The phonon bath is treated as a quantized multi-mode reservoir. In practice this scheme could be realized by the bichromatic driving of an exciton transition between the semiconductor ground state and single exciton state of an InAs/GaAs quantum dot.
Typical parameters of experimental samples of quantum dots~\cite{vc07,rg10} are shapes with heights of $3-5$ nm, base diameters of $25-30$ nm, the exciton transition wavelength $\lambda_{0}=950$ nm ($\omega_{0}/2\pi = 300$ THz).

The total Hamiltonian of the system can be written in the form
\begin{align}
H = H_{0} +H_{1} + H_{2} ,\label{H1}
\end{align}
where $H_{0}$ is the Hamiltonian of the phonon field (setting $\hbar=1$ throughout the paper)
\begin{eqnarray}
H_{0} = \sum_p\omega_{p} b_p^{\dag} b_p ,\label{H2}
\end{eqnarray}
$H_{1}$ is the Hamiltonian of the quantum dot plus the interaction with the bichromatic laser field
\begin{align}
H_{1} &= \omega_{0} S_{z} \nonumber\\
&+\!\left[\Omega_{1}e^{-i(\omega_1 t-\phi_{1}\!)}+\Omega_{2} e^{-i(\omega_2 t-\phi_{2}\!)}\right]S^{+}\!+\!{\rm H.c.} ,\label{H3}
\end{align}
and $H_{2}$ is the interaction Hamiltonian of the quantum dot with the phonon reservoir,
\begin{eqnarray}
H_{2} = \sum_p{g_pS_{z}(b_p+b_p^{\dag})} .\label{H4}
\end{eqnarray}
Here $b_{p}^{\dag}$ and $b_p$ are the creation and annihilation operators of mode $p$ of frequency $\omega_{p}$ of the phonon bath, $S^{+}(S^{-})$ is the raising (lowering) operator and $S_{z}$ is the population difference operator of the quantum dot, and $g_{p}$ is the coupling strength of the mode $p$ of the phonon reservoir to the quantum dot. The parameters $\Omega_{1}$ and $\Omega_{2}$ are the Rabi frequencies between the quantum dot and the components of the laser field, which are given by the product of the atomic transition dipole moment $\vec{\mu}$ and the laser field amplitudes $\vec{\mathcal{E}}_{1}$ and $\vec{\mathcal{E}}_{2}$, respectively.

To remove the fast oscillating terms in Eq.~(\ref{H3}), we transform the Hamiltonian into a frame rotating with the frequency $\omega_{0}$ and obtain
\begin{align}
\tilde{H}_{1} = \left(\Omega_{1}e^{-i(\Delta_{1}t-\phi_1)}
+\Omega_{2} e^{i(\Delta_{2}t+\phi_2)}\right)S^{+} + {\rm H.c.} \label{H5}
\end{align}

We now derive the master equation for the reduced density operator $\rho$ of the quantum dot subject of the driving bichromatic field and the low frequency phonon reservoir. In the treatment, we derive the effective interaction Hamiltonian between the driven QD and the phonon bath, and the derivation closely follows the approach previously used in Refs.~\cite{ah04,mn10,rh11,uw13}.

We assume that the bichromatic field is weak so that the dynamics of the QD are mostly affected by the interaction with the phonon reservoir. In order to analyze the effect of the phonon reservoir on the QD, we define an unitary operator
\begin{eqnarray}
U = i\sum_{p}\frac{g_{p}}{\omega_{p}}\left(b_{p}^{\dag} - b_{p}\right)S_{z} ,\label{H6}
\end{eqnarray}
and make the unitary transformation of the Hamiltonian of the system. Hence, we obtain
\begin{align}
H_{T} = e^{-iU} He^{iU} = H_{R} + H_{I} ,\label{H7}
\end{align}
where
\begin{align}
H_{R} &= e^{-iU} (H_{0} +H_{2}) e^{iU}\nonumber\\
& = \sum_{p} \hbar\omega_{p}b_{p}^{\dag} b_{p} -\sum_{p}\!\frac{g_{p}^{2}}{4\omega_{p}} ,\label{H8}
\end{align}
and
\begin{align}
H_{I} &= e^{-iU}\tilde{H}_{1}e^{iU} = \left[\Omega (t)S^{+} +{\rm H.c.}\right] \nonumber\\
&+ \left[\sum_{p}\frac{g_{p}\Omega(t)}{\omega_p}\left(b_p^{\dag}-b_p\right)S^{+} + {\rm H.c.}\right] + \ldots ,\label{H9}
\end{align}
in which
\begin{align}
\Omega (t) = \left(\tilde{\Omega}_{1} e^{-i\Delta_1 t}+\tilde{\Omega}_2 e^{i\Delta_2 t}\right) ,\label{Omega}
\end{align}
with $\tilde{\Omega}_{i}=\langle B\rangle \Omega_{i}\exp(i\phi_{i})$, and
\begin{align}
\langle B\rangle = \exp\left[-\frac{1}{2}\sum_{p}(g_p/ \omega_p)^2(2\bar{n}_p+1)\right] .\label{Bmean}
\end{align}
Here, $\Omega (t)$ is the total time-dependent Rabi frequency of the driving laser and $\bar n_p\equiv\langle b_p^{\dagger}b_p\rangle=[\exp(\omega_p/k_BT_p)-1]^{-1}$ is the average occupation phonon number of a mode $p$, where $k_B$ is the Boltzmann constant and $T_p$ corresponds to the temperature of the reservoir. The first term in Eq.~(\ref{H8}) represents the energy of the phonon reservoir, while the second term represents a shift of the energy levels of the QD due to the interaction with the phonon reservoir. The shift is known in the literature as the Lamb shift. This term is usually considered to be absorbed into the atomic transition frequency and is not included explicitly in the dynamics of the system. Therefore, the Hamiltonian (\ref{H8}) can be simply considered as the energy of the phonon reservoir.

The first term on the right-hand side of Eq.~(\ref{H9}) contains the interaction of the QD with the driving laser field. The second term represents the interaction of the QD with the phonon reservoir. It is in a form of the electric dipole interaction in which the phonon reservoir couples to the dipole transition of the QD.
In the derivation of Eq.~(\ref{H9}), we have performed a Taylor expansion and have kept only the terms up to first-order in $g_{p}$. With the higher-order terms ignored, we simply limit the interaction of the QD with the phonon reservoir to one-phonon processes only.

It is worthwhile noting at this point that the main result of the unitary transformation of the Hamiltonian of the system is the coupling of the phonon reservoir to the atomic dipole moment. Thus, with the help of the driving laser, the low frequency phonon reservoir effectively couples to the atomic dipole transition~$\ket g \leftrightarrow \ket e$.

We may transform the Hamiltonian (\ref{H7}) into the interaction picture with the unitary operator $U(t)=e^{iH_{R} t}$, and find
\begin{align}
\tilde{H}_{T} = e^{-iH_{R}t} H_{T}e^{iH_{R}t} = V_{L}(t) + V_{R}(t) ,\label{H11}
\end{align}
where $V_{L}(t)$ is the interaction of the laser with the QD, and
\begin{align}
V_{R}(t) = \sum_p\frac{g_{p}\Omega(t)}{\omega_p}\left(b_p^{\dag} e^{i\omega_p t}-b_p e^{-i\omega_p t}\right) \!S^{+}\!+\!{\rm H.c.} \label{H12}
\end{align}
is the interaction of the QD with the phonon reservoir. The interaction $V_{R}(t)$ can be written explicitly as
\begin{align}
V_{R}(t) &= \sum_p\frac{g_{p}}{\omega_p}\left\{\left[b_{p}^{\dag}\left( \tilde{\Omega}_{1}e^{i(\omega_p-\Delta) t} +\tilde{\Omega}_{2}e^{i(\omega_{p}+\Delta) t}\right)\right.\right. \nonumber\\
&\left.\left. -\, b_{p}\left(\tilde{\Omega}_{1} e^{-i(\omega_{p}+\Delta) t}
+ \tilde{\Omega}_{2}e^{-i(\omega_{p}-\Delta) t}\right)\!\right]\!S^{+}\!+\!{\rm H.c.}\right\} ,\label{H13}
\end{align}
where we have assumed that the components of the bichromatic field are equally detuned from the atomic transition frequency, i.e.  $\Delta_{1}=\Delta_{2}\equiv \Delta$.

We see from Eq.~(\ref{H13}) that the interaction contains terms which oscillate at frequencies $\omega_{p}-\Delta$ and $\omega_{p}+\Delta$. If the density of modes of the phonon reservoir is large only in the vicinity of the laser field detuning $\Delta$, then $\omega_{p}\approx \Delta$. In such a case,  the interaction Hamiltonian reduces to resonant, non-oscillating terms, and nonresonant terms oscillating at frequency $2\Delta$. We can make the rotating-wave approximation in which the resonant terms play a dominant role whereas the nonresonant terms make much smaller contributions and can be omitted. The interaction Hamiltonian $V_{R}(t)$ then simplifies~to
\begin{align}
V_{R}(t) =&\, \sum_p\frac{g_{p}}{\omega_p}\left\{\left[\tilde{\Omega}_{1}b^{\dag}_{p}e^{i(\omega_{p}-\Delta) t}\right.\right. \nonumber\\
&\left.\left. - \tilde{\Omega}_{2}b_{p}e^{-i(\omega_{p}-\Delta) t} \right]\!S^{+} + {\rm H.c.}\right\} .\label{H14}
\end{align}

Having derived the effective interaction Hamiltonian of the driven QD with the phonon reservoir, we now turn to the derivation of the master equation for the reduced density operator of the quantum dot
\begin{align}
\rho(t) = {\rm Tr}_{F}W(t) ,
\end{align}
where $W(t)$ is the density operator of the total system, the QD plus the phonon bath. We choose an initial state with no correlations between the QD and the phonon bath modes
\begin{align}
W(0) = \rho_{F}(0)\otimes \rho(0) ,
\end{align}
and specify the phonon bath as a vacuum thermal bath with the following correlations
\begin{align}
\langle b_{p}\rangle &= \langle b^{\dag}_{p}\rangle = 0 ,\quad \langle b^{\dag}_{p}b_{p^{\prime}}\rangle = \bar{n}\delta(p-p^{\prime}) ,\nonumber\\
\langle b_{p}b^{\dag}_{p^{\prime}}\rangle &= (\bar{n}+1)\delta(p-p^{\prime}) ,
\end{align}
where $\bar{n}$ is the average number of phonons.

After tracing over the phonon bath operators, and using the standard Born-Markov approximations, we arrive at the master equation
\begin{eqnarray}
\frac{\partial}{\partial t}{\rho} = -i[V_{L}(t),\rho] +\mathcal{L}_{b}\rho + \mathcal{L}_{p}\rho ,\label{H15}
\end{eqnarray}
in which
\begin{align}
V_{L}(t) = \Omega (t)S^{+} +{\rm H.c.} \label{H16}
\end{align}
is the interaction of the quantum dot with the bichromatic field,
\begin{align}
\mathcal{L}_{b}\rho = \frac{1}{2}\Gamma\left(2S_{-}\rho S_{+}-S_{+}S_{-}\rho-\rho S_{+}S_{-}\right) \label{H17}
\end{align}
represents the damping of the quantum dot at the rate $\Gamma$ by spontaneous emission to vacuum radiation modes, other than the phonon modes, and
\begin{align}
\mathcal{L}_{p}\rho &= \gamma_{s}(2S^{-}\rho S^{+}-S^{+}S^{-}\rho-\rho S^{+}S^{-}) \nonumber\\
&+\gamma_{n}(2S^{+}\rho S^{-}-S^{-}S^{+}\rho-\rho S^{-}S^{+}) \nonumber\\
&-\gamma_{m}\left(2S^{+}\rho S^{+} e^{2i\phi}+2S^{-}\rho S^{-} e^{-2i\phi}\right) ,\label{H18}
\end{align}
represents the damping of the quantum dot by a reservoir mediated by the phonon bath modes and the bichromatic field. Here, the parameters are defined as
\begin{align}
\gamma_{s} &= \gamma_1 \bar{n}+\gamma_2 (\bar{n}+1) ,\quad
\gamma_{n} = \gamma_{1}(\bar{n}+1)+\gamma_2 \bar{n} ,\nonumber\\
\gamma_{m} &= (2 \bar{n}+1)\sqrt{\gamma_1 \gamma_2} ,\quad  2\phi =\phi_{1}+\phi_{2} ,\label{H19}
\end{align}
with
\begin{align}
\gamma_i = 2\pi \tilde{\Omega}_{i}^2 \sum_{p}\left(\frac{g_p}{\omega_p}\right)^2\delta(\omega_p-\Delta) ,\  i=1,2 .\label{H20}
\end{align}
The assumption of the Born approximation, valid to second orders in the quantum dot environment coupling strengths, $|g_{p}|^{2}$ and $|f_{k}|^{2}$, where $f_{k}$ is the coupling strength of the $k$th radiation mode to the quantum dot, restricts the master equation to weak-system reservoir coupling regimes. The assumption of the Markov approximation restricts the master equation to times $t$ longer than the time $\Delta t$ required for a phonon to traverse the quantum dot, $t\gg \Delta t =l/u$, where $l$ is the size of a quantum dot and $u$ is the speed of sound. Based on the typical sizes of experimental samples of an InAs/GaAs quantum dot of $l=5$ nm, $\Delta t \approx 1$ ps. For the Markov approximation to be valid, the time $\Delta t$ should be shorter than any relaxation time in the system. In practice this may well be a reasonable assumption. For example, the radiative recombination  time of the exciton, determined by $1/\Gamma$, is usually  $500 - 800$ ps~\cite{mn13,ua11}.

\begin{figure}[h]
\centering\includegraphics[width=.9\columnwidth,keepaspectratio,clip]{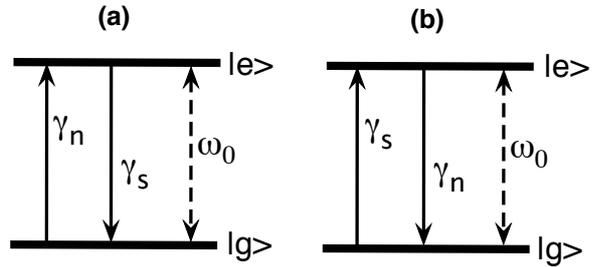}
\caption{Illustration of the role of the parameters $\gamma_{s}$ and $\gamma_{n}$ in the dynamics of the quantum dot for (a) $\gamma_{2}>\gamma_{1}$ and (b) $\gamma_{1}>\gamma_{2}$.}
\label{bfig2}
\end{figure}

The Liouvillian (\ref{H18}) has a structure analogous to the damping of a two-level system by a squeezed reservoir. The parameters $\gamma_{s}$ and $\gamma_{n}$ correspond to incoherent damping and incoherent pumping rates, respectively, and $\gamma_{m}$ corresponds to the strength of two-photon correlations. A close look at the parameters in Eq.~(\ref{H19}) reveals that not always $\gamma_{s}$ constitutes the incoherent rate at which a population is damped by the reservoir. We can have $\gamma_{s}>\gamma_{n}$ as well as $\gamma_{n}>\gamma_{s}$. Which of these takes place depends principally on whether $\gamma_{2}>\gamma_{1}$ or $\gamma_{1}>\gamma_{2}$. For $\gamma_{2}>\gamma_{1}$, we have $\gamma_{s}>\gamma_{n}$. In this case, $\gamma_{s}$ can be viewed as the incoherent damping rate. Otherwise, when $\gamma_{2}<\gamma_{1}$, we have $\gamma_{n}>\gamma_{s}$ that the incoherent pumping rate exceeds the damping rate. In this case, the reservoir is formed from a bath of inverted harmonic oscillators, and the rate of transferring the population from the ground state $\ket 1$ to the upper state $\ket 2$ is larger than the rate of transferring the population from $\ket 2$ to $\ket 1$. It is easy to see from Eq.~(\ref{H19}) that the condition of $\gamma_{1}>\gamma_{2}$ can be achieved when the Rabi frequency $\tilde{\Omega}_{1}$ of the bichromatic field component tuned above the resonance exceeds the Rabi frequency $\tilde{\Omega}_{2}$ of the component tune below the resonance. Figure~\ref{bfig2} illustrates the role of the parameters $\gamma_{s}$ and $\gamma_{n}$ in the dynamics of the quantum dot. It is seen that the roles of the parameters reverse when $\gamma_{2}>\gamma_{1}$ reverses to $\gamma_{1}>\gamma_{2}$.

Apart from the damping by the squeezed reservoir there is also a contribution from the damping by the radiation field modes, Eq.~(\ref{H17}).
If the phonon bath modes occupy all modes to which the quantum dot is coupled, $\Gamma \sim |f_{k}|^{2} =0$, then the quantum dot is damped solely by the squeezed reservoir. However, if there is a small fraction of modes not occupied by the phonon bath modes, $\Gamma\neq 0$, then the quantum dot is damped by both the squeezed reservoir and the "unsqueezed" radiation modes.

To investigate how efficient the phonon bath together with the bichromatic field is in the creation of a squeezing type reservoir to the quantum dot, we compare the master equation~(\ref{H18}) with the equation when a two-level system is illuminated by a squeezed vacuum field produced by an external squeezing source, such as an optical parametric oscillator~\cite{wk86,wx87,mw13}. The squeezed vacuum field is characterized by the correlation functions~\cite{ga86,cl87,pa90,df99,df04}
\begin{align}
\langle a_{k}a_{k^{\prime}}^{\dag}\rangle &= (N+1)\delta_{k,k^{\prime}} ,\nonumber\\
\langle a_{k}^{\dag}a_{k^{\prime}}\rangle &= N\delta_{k,k^{\prime}} ,\nonumber\\
\langle a_{k}a_{k^{\prime}}\rangle &= |M|e^{-2i\Phi}\delta_{2k_{s}-k,k^{\prime}} ,\nonumber\\
\langle a_{k}^{\dag}a_{k^{\prime}}^{\dag}\rangle &= |M|e^{2i\Phi}\delta_{2k_{s}-k,k^{\prime}} ,\label{H20a}
\end{align}
where $a_{k} (a_{k}^{\dag})$ is the annihilation (creation) operator for mode $k$ of the squeezed field, $N$ is the number of photons in the field, the parameter $|M|$ determines the degree of two-photon (squeezing) correlations between modes symmetrically located about the squeezing carrier mode $2k_{s}$, and $\Phi$ is the phase of the field. The parameter $|M|$ may fall into one of the two separate regions
\begin{align}
|M|-N< 0 \quad {\rm or}\quad N<|M| \leq \sqrt{N(N+1)} .\label{H22}
\end{align}
If $|M|$ falls into the region of $|M|<N$, the field corresponds to the so-called classically squeezed field in the sense that fluctuations in one of the quadratures of the field amplitudes are reduced but not below the shot-noise level. If $|M|$ falls into the region of $N<|M| \leq \sqrt{N(N+1)}$, the field is then a quantum squeezed field in the sense that the fluctuations of one of the quadratures are suppressed below the shot-noise level. The equality $|M| = \sqrt{N(N+1)}$ corresponds to maximal correlations, an ideal squeezed field. Thus, there are lower and upper limits, $|M|=N$ and $|M|=\sqrt{N(N+1)}$, respectively, for the quantum correlations of the squeezed field.

The interaction of a two-level system with a reservoir characterized by the correlation functions (\ref{H20a}) leads to the following master equation~\cite{ga86,cl87,pa90,df99,df04} 
\begin{align}
\mathcal{L}_{p}\rho &= \frac{1}{2}\gamma\left(N+1\right)(2S^{-}\rho S^{+}-S^{+}S^{-}\rho-\rho S^{+}S^{-})\nonumber\\
&+\frac{1}{2}\gamma N(2S^{+}\rho S^{-}-S^{-}S^{+}\rho-\rho S^{-}S^{+})\nonumber\\
&-\frac{1}{2}\gamma |M|\left(2S^{+}\rho S^{+} e^{2i\Phi}+2S^{-}\rho S^{-} e^{-2i\Phi}\right) ,\label{H21}
\end{align}
where $\gamma$ is the spontaneous emission rate of the atomic transition.

Matching coefficients in Eqs.~(\ref{H18}) and (\ref{H21}), we find that for~$\gamma_{2}>\gamma_{1}$:
\begin{align}
\gamma_{s} -\gamma_{n} \rightarrow \frac{1}{2}\gamma ,\quad
\gamma_{n} \rightarrow \frac{1}{2}\gamma N ,\quad
\gamma_{m} \rightarrow \frac{1}{2}\gamma |M| .\label{H23}
\end{align}
Similarly, for $\gamma_{1}>\gamma_{2}$:
\begin{align}
\gamma_{n} -\gamma_{s} \rightarrow \frac{1}{2}\gamma_{I} ,\quad
\gamma_{s} \rightarrow \frac{1}{2}\gamma_{I} N ,\quad
\gamma_{m} \rightarrow \frac{1}{2}\gamma_{I} |M| ,\label{H24}
\end{align}
where the subscript $"I"$ stands for inverted harmonic oscillator. Thus, the effective squeezing type reservoir can be described by field operators $b_{k}$ and $b_{k}^{\dag}$ satisfying the following correlation functions $(\gamma_{2}>\gamma_{1})$:
\begin{align}
\langle b_{k}b_{k^{\prime}}^{\dag}\rangle &= \frac{\gamma_{s}}{\gamma_{2}-\gamma_{1}}\delta_{k,k^{\prime}} ,\nonumber\\
\langle b_{k}^{\dag}b_{k^{\prime}}\rangle &= \frac{\gamma_{n}}{\gamma_{2}-\gamma_{1}}\delta_{k,k^{\prime}} ,\nonumber\\
\langle b_{k}b_{k^{\prime}}\rangle &= \frac{\gamma_{m}}{\gamma_{2}-\gamma_{1}}e^{-2i\phi}\delta_{2k_{s}-k,k^{\prime}} ,\nonumber\\
\langle b_{k}^{\dag}b_{k^{\prime}}^{\dag}\rangle &= \frac{\gamma_{m}}{\gamma_{2}-\gamma_{1}}e^{2i\phi}\delta_{2k_{s}-k,k^{\prime}} .\label{H20b}
\end{align}

As we have mentioned above, there are lower and upper limits for the quantum correlations of the squeezed field. If we evaluate the lower limit $|M|-N$ according to Eq.~(\ref{H23}), we find the result
\begin{align}
|M|-N = \frac{\sqrt{\gamma_{1}}-\bar{n}\left(\sqrt{\gamma_{2}}-\sqrt{\gamma_{1}}\right)}{\sqrt{\gamma_{1}}+\sqrt{\gamma_{2}}}  ,\label{H25}
\end{align}
and
\begin{align}
|M|-N = \frac{\sqrt{\gamma_{2}}-\bar{n}\left(\sqrt{\gamma_{1}}-\sqrt{\gamma_{2}}\right)}{\sqrt{\gamma_{1}}+\sqrt{\gamma_{2}}}  ,\label{H26}
\end{align}
if we evaluate the limit according to Eq.~(\ref{H24}). It is seen from Eqs.~(\ref{H25}) and (\ref{H26}) that for $|M|$ to fall into the region of quantum squeezing, $|M|-N>0$, it is necessary that
\begin{align}
\bar{n} &< \frac{\sqrt{\gamma_{1}}}{\sqrt{\gamma_{2}}-\sqrt{\gamma_{1}}} ,\quad \gamma_{2}>\gamma_{1} ,\nonumber\\
\bar{n} &< \frac{\sqrt{\gamma_{2}}}{\sqrt{\gamma_{1}}-\sqrt{\gamma_{2}}} ,\quad \gamma_{1}>\gamma_{2} ,\label{H27}
\end{align}
and if these conditions hold, then $N<|M| \leq \sqrt{N(N+1)}$. Thus, the phonon bath with the help of the bichromatic field creates a squeezed reservoir which can be unique to the quantum field.

If we evaluate the upper limit for the correlations, $|M|^{2}-N(N+1) =0$, which determines the maximal two-photon (squeezing) correlations in the field, we find
\begin{align}
|M|^{2} -N(N+1) = -\bar{n}(\bar{n}+1) ,\label{H28}
\end{align}
for both $\gamma_{2}>\gamma_{1}$ and $\gamma_{1}>\gamma_{2}$ cases.
We see that maximal correlations are achieved only at $\bar{n}=0$. For $\bar{n}\neq 0$ the reservoir appears as an imperfectly squeezed reservoir with the correlations decreasing with an increasing~$\bar{n}$. It is interesting to note that the upper limit for the correlations is independent of $\gamma_{1}$ and~$\gamma_{2}$, whereas the lower limit, as seen from Eqs.~(\ref{H25}) and (\ref{H26}), varies with $\gamma_{1}$ and~$\gamma_{2}$.
\begin{figure}[h]
\centering\includegraphics[width=\columnwidth,keepaspectratio,clip]{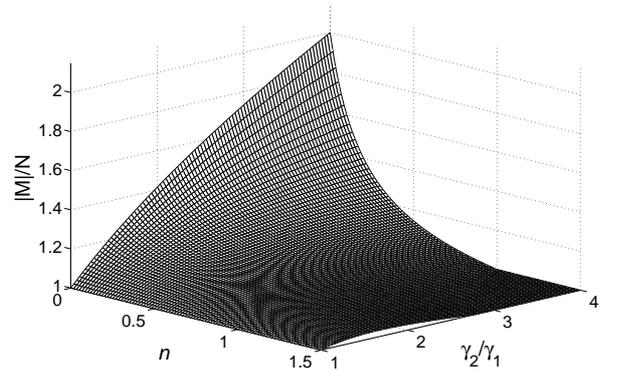}
\caption{The ratio $|M|/N$ plotted as a function of $\bar{n}$ and $\gamma_{2}/\gamma_{1}$. The quantum nature of the correlations $(|M|/N>1)$ occurs for $\bar{n}< 1/(\sqrt{\gamma_{2}/\gamma_{1}} -1)$. }
\label{bfig3}
\end{figure}

Figure~\ref{bfig3} shows the ratio $|M|/N$ as a function of $\bar{n}$ and $\gamma_{2}/\gamma_{1}$ for the case of $\gamma_{2}>\gamma_{1}$. Values $|M|/N>1$ signal the quantum nature of the correlations. The correlations increase with an increasing $\gamma_{2}/\gamma_{1}$ and can reach large values but the largest increase of the correlations above the classical limit occurs for weak squeezed fields, i.e., for large $\gamma_{2}/\gamma_{1}$ at which $N$ is small. The correlations are very sensitive to~$\bar{n}$. For not too large $\gamma_{2}/\gamma_{1}$, the ratio decreases slowly with $\bar{n}$ so that the quantum nature of the correlations persists at large $\bar{n}$. However, for large $\gamma_{2}/\gamma_{1}$, the decrease of the ratio with $\bar{n}$ is considerably more rapid than it is for small $\gamma_{2}/\gamma_{1}$ that the quantum correlations decrease rapidly with $\bar{n}$. Beyond $\bar{n}=1/(\sqrt{\gamma_{2}/\gamma_{1}} -1)$, the ratio falls below the lower limit for quantum correlations.

The case $\gamma_{1}=\gamma_{2}$ has to be treated separately. In the limit $\gamma_{1}=\gamma_{2}\equiv\gamma_{0}$, the coefficients in Eq.~(\ref{H18}) are
\begin{align}
\gamma_{s} = \gamma_{n} =\gamma_{m} = \left(2\bar{n}+1\right)\gamma_{0} .\label{H29}
\end{align}
When compared with the coefficients of Eq.~(\ref{H21}), we find
\begin{align}
\gamma_{s} &\rightarrow \frac{1}{2}\gamma (N+1) ,\quad \gamma_{n}\rightarrow \frac{1}{2}\gamma N ,\nonumber\\
\gamma_{m} &\rightarrow \frac{1}{2}\gamma |M| =\sqrt{\gamma_{n}\gamma_{s}}\rightarrow \frac{1}{2}\gamma\sqrt{N(N+1)} .\label{H30}
\end{align}
Since $\gamma_{s}=\gamma_{n}$, this limit can be regarded as corresponding to a very strong squeezed field, $N\gg 1$, with maximal correlations $|M|=\sqrt{N(N+1)}$. We see that in this case the upper limit of the squeezing correlations is achieved for any values of $\bar{n}$.

\section{Dynamics of the quantum dot}\label{sec3}

Let us now apply these considerations explicitly to the dynamics of the quantum dot interacting with the engineered squeezed reservoir. The dynamical response of the quantum dot interacting with the squeezed reservoir is best described in terms of the expectation values of the dipole components which obey the following optical Bloch equations
\begin{align}
\dot{\av{S_{x}}} =& -\left(\frac{1}{2}\Gamma +\gamma_{s}+\gamma_{n}+2\gamma_{m}\cos2\phi\right)\av{S_{x}} \nonumber\\
&+ 2\gamma_{m}\sin2\phi \av{S_{y}} ,\nonumber\\
\dot{\av{S_{y}}} =& -\left(\frac{1}{2}\Gamma +\gamma_{s}+\gamma_{n}-2\gamma_{m}\cos2\phi\right)\av{S_{y}} \nonumber\\
&+ 2\gamma_{m}\sin2\phi \av{S_{x}} ,\nonumber\\
\dot{\av{S_{z}}} =& -\left(\gamma_{s}-\gamma_{n} +\frac{1}{2}\Gamma\right) -2\left(\gamma_{s}+\gamma_{n}+\frac{1}{2}\Gamma\right)\av{S_{z}} ,\label{H31}
\end{align}
where $S_{x}=(S^{-}+S^{+})/2$ and $S_{y}=i(S^{-}-S^{+})/2$ are the dipole polarization components.
Te effect of squeezing is best seen through the quadrature phase components
\begin{align}
S_{\phi}(t) &= S_{x}(t)\sin\phi +S_{y}(t)\cos\phi ,\nonumber\\
S_{\phi+\frac{\pi}{2}}(t) &= S_{x}(t)\cos\phi -S_{y}(t)\sin\phi .\label{H32}
\end{align}

When Eq.~(\ref{H31}) is integrated, we obtain
\begin{align}
\av{S_{\phi}(t)} &= \av{S_{\phi}(0)}e^{-\left(\frac{1}{2}\Gamma +\gamma_{s}+\gamma_{n}-2\gamma_{m}\right)t} ,\nonumber\\
\av{S_{\phi+\frac{\pi}{2}}(t)} &= \av{S_{\phi+\frac{\pi}{2}}(0)} e^{-\left(\frac{1}{2}\Gamma +\gamma_{s}+\gamma_{n}+2\gamma_{m}\right)t} ,\nonumber\\
\av{S_{z}(t)} &= -\frac{\gamma_{s}-\gamma_{n}+\frac{1}{2}\Gamma}{2\left(\gamma_{s}+\gamma_{n}+\frac{1}{2}\Gamma\right)} \nonumber\\
&+\!\left[\!\av{S_{z}(0)} +\frac{\gamma_{s}-\gamma_{n}+\frac{1}{2}\Gamma}{2\!\left(\gamma_{s}\!+\!\gamma_{n}\!+\!\frac{1}{2}\Gamma\right)}\right]\!e^{-2\left(\frac{1}{2}\Gamma +\gamma_{s}+\gamma_{n}\right)t} .\label{H33}
\end{align}
The components display simple exponential decays, the component $\av{S_{\phi}(t)}$ is damped at a reduced rate $\gamma_{\phi}= \frac{1}{2}\Gamma +\gamma_{s}+\gamma_{n}-2\gamma_{m}$, while $\av{S_{\phi+\frac{\pi}{2}}(t)}$ is damped at an enhanced rate $\gamma_{\phi+\frac{\pi}{2}}=\frac{1}{2}\Gamma +\gamma_{s}+\gamma_{n}+2\gamma_{m}$. The population inversion $\av{S_{z}(t)}$ decays to a steady-state value which depends strongly on the relation between $\gamma_{s}$ and $\gamma_{n}$.

The decay rates depend also on $\Gamma$. Because it is precisely the effect of the phonon bath on the dynamics of the quantum dot that interests us most here, in what follows, we shall assume $\gamma_{s},\gamma_{n}\gg\Gamma$ and set $\Gamma =0$. This is justified if one notices from Eq.~(\ref{H20}) that $\gamma_{s}$ and $\gamma_{n}$ increase with an increasing $\tilde{\Omega}_{i}$. Thus, we may increase the Rabi frequencies of the bichromatic field such that $\gamma_{s,n}\gg \Gamma$. This situation can be achieved in current experiments since the radiative lifetime $500 - 800$ ps corresponds to $\Gamma \sim 1.2$ GHz. Using the definition $J(\omega) =\sum_{p}g_{p}^{2}\delta(\omega -\omega_{p})$, which is equivalent to $J(\omega)=\alpha \omega^{3}\exp[-(\omega/\omega_{c})^{2}]$, where $\omega_{c}=\sqrt{2}u/l$ is the cutoff frequency~\cite{na08}, the damping parameters $\gamma_{i}$ can be estimated by writing~\cite{wh13}
\begin{align}
\gamma_i = 2\pi \tilde{\Omega}_{i}^2 \sum_{p}\left(\frac{g_p}{\omega_p}\right)^2\delta(\omega_p-\Delta) = 2\pi \tilde{\Omega}_{i}^2 \alpha \Delta .
\end{align}
For the Rabi frequencies of the bichromatic field we choose the average value $\Omega_{i}= 70$ GHz~\cite{wh13}. For the phonon bath we assume $\bar{n}=0.5$ and a temperature $T= 2.35$ K, and take $\omega_{c} =1500$ GHz~\cite{na08}. These give $\Delta (=\omega_{p}) = 490$ GHz. Taking $\alpha = 2.535\times 10^{-7}$ (GHz)$^{-2}$~\cite{mn10,wh13}, the damping parameters $\gamma_{i}$ are then $\gamma_{i}= 4$ GHz, which are larger than $\Gamma =1.2$ GHz.

As discussed in Sec.~\ref{sec2}, the manner in which the squeezed reservoir can affect the dynamics of the quantum dot depends on the relation between $\gamma_{1}$ and $\gamma_{2}$. There are three distinct regimes of the parameters at which the effective squeezed reservoir can have significantly different properties, (A.) $\gamma_{2}>\gamma_{1}$, (B.) $\gamma_{1}> \gamma_{2}$, and (C.) $\gamma_{1}=\gamma_{2}$.

\subsection{The case, $\gamma_{2}>\gamma_{1}$}

This limit corresponds to $\gamma_{s}>\gamma_{n}$ that the effective reservoir is formed from a phonon bath of the ordinary harmonic oscillators.
As shown in Sec.~\ref{sec2}, there is in this case a direct correspondence between the engineered squeezed reservoir and that produced by an external source of the squeezed vacuum field.
Therefore, the effects of the engineered squeezed reservoir on the dynamics of the quantum dot are expected to be analogous to those which are well known for a two-level atom damped by a squeezed vacuum field produced by an external source. However, there are some subtle differences. For example, if an external source produces a squeezed field with maximal correlations, $|M|=\sqrt{N(N+1)}$, the correlations remain maximal independent of the value of $N$. However, in the squeezed reservoir engineered from a phonon bath, an increase of the number of phonons $\bar{n}$ is accompanied by a decrease of the two-photon correlations $|M|$, as seen from Eq.~(\ref{H28}). As a consequence, ideally squeezed reservoir at $\bar{n}=0$ becomes an imperfectly squeezed reservoir when $\bar{n}\neq 0$. Thus, we lose the option of having an ideally squeezed reservoir when $\bar{n}\neq 0$.
Under this circumstance, many effects unique to the quantum nature of the squeezed field may not be observed due to the presence of thermal phonons.

In order to determine the range of the parameters, $\bar{n}$ in particular, at which the effects unique to the quantum nature of the squeezed field still could be observed in the presence of thermal phonons, we write $N=N_{s}+N_{b}$ such that $N_{s}(N_{s}+1)=|M|^{2}$. Thus, $N_{s}$ corresponds to the number of phonons in the maximally squeezed field and $N_{b}$ is the number of excess phonons, which are not correlated, and therefore correspond to a thermal (background) field. When this division of $N$ is applied to Eq.~(\ref{H18}), it is straightforward to find from Eq.~(\ref{H19}) that
\begin{align}
N_{s} &= \sqrt{4\bar{n}(\bar{n}+1)u^{2} +w^{2}} -\frac{1}{2},\nonumber\\
N_{b} &= (2\bar{n}+1)w -\sqrt{4\bar{n}(\bar{n}+1)u^{2}+w^{2}} ,\label{H34}
\end{align}
where $u=2\sqrt{\gamma_{1}\gamma_{2}}/\gamma$ and $w=(\gamma_{1}+\gamma_{2})/\gamma$.
The Liouvillian~(\ref{H18}) can then be written in the form
\begin{align}
\mathcal{L}_{p}\rho &= \frac{1}{2}\gamma (2\Upsilon\rho \Upsilon^{\dag}-\Upsilon^{\dag}\Upsilon\rho-\rho \Upsilon^{\dag}\Upsilon) \nonumber\\
&+\frac{1}{2}\gamma N_{b}(2S^{-}\rho S^{+}-S^{+}S^{-}\rho-\rho S^{+}S^{-}) \nonumber\\
&+\frac{1}{2}\gamma N_{b}(2S^{+}\rho S^{-}-S^{-}S^{+}\rho-\rho S^{-}S^{+}) ,\label{H35}
\end{align}
where
\begin{align}
\Upsilon = \sqrt{N_{s}+1}S^{-}e^{-i\phi} +\sqrt{N_{s}}S^{+}e^{i\phi} .\label{H36}
\end{align}
Thus, the Liouvillian (\ref{H18}) describing the damping of the quantum dot by an imperfectly squeezed reservoir has been divided into two parts, one describing damping by the maximally squeezed reservoir, the first term in Eq.~(\ref{H35}), and the other describing damping by the background thermal reservoir, the second and third terms in (\ref{H35}). In other words, the interaction of the quantum dot with an imperfect squeezed reservoir may be viewed as the interaction with two separate reservoirs, a maximally correlated squeezed reservoir and a thermal reservoir.
Which reservoir dominates in the interaction depends on the ratio $N_{b}/N_{s}$. If $N_{s}$ is much larger than $N_{b}$, then significant effects of the squeezed reservoir should be observable. For example, both $N_{b}$ and $N_{s}$ contribute to the damping rate of the quantum dot. Therefore, the reduction of $\gamma_{\phi}$ below the standard quantum limit is possible only for $N_{b}<|M|-N_{s}$. Viewed as a function of $\bar{n}$, the inequality $N_{b}<|M|-N_{s}$ becomes $\bar{n}<1/(\sqrt{\gamma_{2}/\gamma_{1}} -1)$ which is, as one could expect, the same as the condition (\ref{H27}) for quantum squeezing.
\begin{figure}[h]
\centering\includegraphics[width=\columnwidth,keepaspectratio,clip]{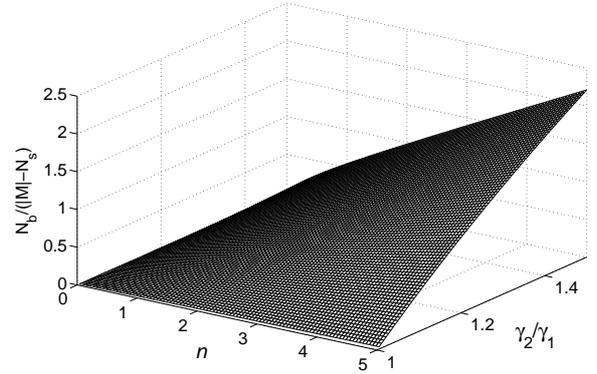}
\caption{The ratio of the number of the background thermal phonons~$N_{b}$ to the degree of quantum squeezing correlations $|M|-N_{s}$ plotted as a function of $\bar{n}$ and $\gamma_{2}/\gamma_{1}$. }
\label{bfig4}
\end{figure}

Figure~\ref{bfig4} shows the ratio $N_{b}/(|M|-N_{s})$ as a function of $\bar{n}$ and $\gamma_{2}/\gamma_{1}$.
The grow of the ratio with $\bar{n}$ depends strongly on $\gamma_{2}/\gamma_{1}$ and the effect of increasing $\gamma_{2}/\gamma_{1}$ is to decrease the range of $\bar{n}$ over which the ratio is smaller than 1.

\subsection{The case, $\gamma_{1}>\gamma_{2}$}

The exchange $\gamma_{1}\leftrightarrow \gamma_{2}$ leads to an exchange $\gamma_{s}\leftrightarrow\gamma_{n}$. Therefore, the damping rates of the quadrature components in Eq.~(\ref{H31}) are formally identical with the corresponding damping rates for $\gamma_{2}>\gamma_{1}$, and the interpretation of the properties of the Liouvillian (\ref{H18}) follows the same lines as for $\gamma_{2}>\gamma_{1}$.

There is, however, an important difference in the evolution of the average inversion $\av{S_{z}(t)}$.
In the parameter regime, $\gamma_{n}>\gamma_{s}$, an incoherent excitation of the quantum dot is more likely to be followed with a further excitation than with a spontaneous decay to the ground level. The net effect of these processes is to accumulate more population in the excited state rather than in the ground state. In physical terms, the system behaves as an inverted harmonic oscillator. It is clearly seen from Eq.~(\ref{H31}) that in the steady-state the population inversion~is
\begin{align}
\av{S_{z}}_{s} = \frac{\gamma_{n}-\gamma_{s}}{2(\gamma_{s}+\gamma_{n})} = \frac{\gamma_{1}-\gamma_{2}}{2(2\bar{n}+1)(\gamma_{1}+\gamma_{2})}  .\label{H37}
\end{align}
Clearly, the population between the bare states of the quantum dots is inverted. The population inversion increases with an increasing $\gamma_{1}/\gamma_{2}$ and for $\gamma_{1}/\gamma_{2}\gg 1$ can reach the total inversion. The effect of the squeezed reservoir on the dynamics of the quantum dot is therefore much more drastic when $\gamma_{1}>\gamma_{2}$ than when $\gamma_{2}>\gamma_{1}$. The result (\ref{H37}) implies that the engineered squeezed reservoir can be employed to maintain a large population inversion necessary for laser generation in a two-level quantum dot.

\subsection{The case, $\gamma_{1}=\gamma_{2}$}

In this limit $\gamma_{s}=\gamma_{n}=\gamma_{m}$, and then the damping rate of the $\av{S_{\phi}(t)}$ component
\begin{align}
\gamma_{\phi} = \gamma_{s}+\gamma_{n}-2\gamma_{m} ,\label{H38}
\end{align}
reduces to zero. Consequently, the decay of $\av{S_{\phi}(t)}$ is completely inhibited that the squeezing of the fluctuations of the~$S_{\phi}$ component is perfect. This shows that by a proper engineering of the coupling of a phonon bath to a quantum dot one can produce a coherent atomic dipole without the accompanying quantum fluctuations and incoherent excitations associated with the presence of phonons. Note that $\gamma_{\phi}=0$ is obtained independent of the number of phonons $\bar{n}$. It follows that $\av{S_{\phi}(t)}$ can be locked at its initial value for an arbitrary long time independent of the temperature of the phonon bath.

On the other hand, the component $S_{\phi+\frac{\pi}{2}}(t)$ decays at the rate $\gamma_{\phi+\frac{\pi}{2}}= 4\left(2\bar{n}+1\right)\gamma_{0}$, which is enhanced and dependent on $\bar{n}$ but is not infinite. Moreover, the two-level transition in the quantum dot becomes saturated in the steady-state, that $\lim_{t\rightarrow\infty}\av{S_{z}(t)}\equiv \av{S_{z}}_{s} =0$. Thus, independent of $\bar{n}$ the population is evenly distributed between the ground and excited levels of the quantum dot. Therefore, the interaction of the quantum dot with the squeezed reservoir (\ref{H18}) offers the possibility of both inhibiting the phase decay and an alignment of the spin vector along the $x$ axis.

These features are significantly different from those produced by the decay of a two-level system to a squeezed reservoir generated by an external source. When the Liouvillian (\ref{H21}) is used instead of (\ref{H18}), one can easily find that the components display the following exponential decays
\begin{align}
\av{S_{\phi}(t)} &= \av{S_{\phi}(0)}e^{-\gamma (\frac{1}{2} + N - |M|)t} ,\nonumber\\
\av{S_{\phi+\frac{\pi}{2}}(t)} &= \av{S_{\phi+\frac{\pi}{2}}(0)} e^{-\gamma(\frac{1}{2}+ N+|M|)t} ,\nonumber\\
\av{S_{z}(t)} &= -\frac{1}{2N+1} \nonumber\\
&+\left[\av{S_{z}(0)} +\frac{1}{2N+1}\right]e^{-\gamma(2N+1)t} .\label{H39}
\end{align}
Clearly, for the inhibition of the decay of the component $\av{S_{\phi}(t)}$ one evidently requires a very strong squeezed field, $N\rightarrow\infty$ at which $N-|M|\rightarrow -\frac{1}{2}$. In this limit, the decay rate of the $\av{S_{\phi+\frac{\pi}{2}}(t)}$ goes to infinity. Moreover, $\av{S_{z}}_{s} <0$ and the population inversion approaches zero only in the limit of $N\rightarrow\infty$.

The physical reason for the changed decay behavior in the engineered squeezed reservoir is most clearly understood by considering the expectation value of the spin vector of the quantum dot and its fluctuations. In the steady-state, we have
\begin{align}
\av{S_{\phi}}_{s} = \av{S_{\phi}(0)} ,\quad \av{S_{\phi+\frac{\pi}{2}}}_{s} = 0 ,\quad \av{S_{z}}_{s} = 0 .\label{H40}
\end{align}
Thus, the expectation value of the spin vector $\av{\vec{S}}$ lies in the $x-y$ plane such that it forms an angle $\phi$ with the $y$ axis
\begin{eqnarray}
     \av{S_{x}}_{s} = S\sin\phi ,\quad  \av{S_{y}}_{s} = S\cos\phi ,\quad \av{S_{z}}_{s} = 0 ,\label{H41}
\end{eqnarray}
where $S=\sqrt{\av{S_{x}}^{2}+\av{S_{y}}^{2}}$ and $\tan\phi = \av{S_{x}}/\av{S_{y}}$.

Assume for simplicity that $\phi=0$.
In this case, the following Heisenberg uncertainty principles are obeyed
\begin{align}
     &\sqrt{\langle (\Delta S_{x})^{2}\rangle \langle (\Delta S_{y})^{2}\rangle } \geq 0 ,\quad x-y\quad  {\rm plane} ,\nonumber\\
     &\sqrt{\langle (\Delta S_{z})^{2}\rangle \langle (\Delta S_{x})^{2}\rangle } \geq \frac{1}{2}|\av{S_{y}}_{s}| ,\quad z-x\quad {\rm plane} ,\nonumber\\
     &\sqrt{\langle (\Delta S_{y})^{2}\rangle \langle (\Delta S_{z})^{2}\rangle } \geq 0  ,\quad y-z\quad {\rm plane} .\label{H42}
\end{align}
The form of the uncertainty  relations resembles very much  that occurring in a planar squeezing situation where one can independently change fluctuations in two quadrature components which lie in the plane of the spin vector~\cite{hp11,hv12,dg15}.

It is not difficult to show that in the case considered here the Liouvillian (\ref{H18}) can be written as
\begin{align}
\mathcal{L}_{p}\rho = 4(2\bar{n}+1)\gamma_{0}\left(2S_{\phi}\rho S_{\phi} -S^{2}_{\phi}\rho -\rho S^{2}_{\phi}\right) ,\label{H43}
\end{align}
from which we see that only the quadrature phase $S_{\phi}$ is coupled to the reservoir.
Thus, we conclude that the case $\gamma_{1}=\gamma_{2}$ corresponds to a quantum-nondemolition type coupling of the quantum dot to the effective squeezed reservoir~\cite{wm85,pc96}.

\section{Stationary state of a driven quantum dot}\label{sec4}

Suppose that in addition to the bichromatic field, which couples the quantum dot to the phonon bath, the dot is subjected to an exciting  laser field of frequency $\omega_{L}$ which is on resonance with the transition frequency of the quantum dot, i.e., detuning $\Delta_{L}=\omega_{L}-\omega_{0}=0$. With the addition of the exciting field, the Bloch equations~(\ref{H31}) take the form
\begin{align}
\dot{\av{S_{x}}} &= -\gamma_{x}\av{S_{x}} ,\nonumber\\
\dot{\av{S_{y}}} &= -\gamma_{y}\av{S_{y}} -\Omega \av{S_{z}} ,\nonumber\\
\dot{\av{S_{z}}} &= -(\gamma_{s}-\gamma_{n}) -\gamma_{z}\av{S_{z}} +\Omega \av{S_{y}} ,\label{H44}
\end{align}
where
\begin{align}
\gamma_{x} &= \gamma_{s}+\gamma_{n}\pm 2\gamma_{m} ,\quad \gamma_{y}= \gamma_{s}+\gamma_{n}\mp 2\gamma_{m} ,\nonumber\\
\gamma_{z} &= 2(\gamma_{s}+\gamma_{n}) ,\label{H45}
\end{align}
and $\Omega$ is the Rabi frequency of the exciting field. In writing Eq.~(\ref{H44}) we have chosen the phase of the laser $\phi_{L}=0$ and have made the choices of the squeezing phase $\phi=0$ and $\phi=\pi/2$ corresponding to the limits of the variation of the damping rates.

Our purpose is to determine the steady-state values of the average values of the spin components. It is clear from Eq.~(\ref{H44}) that the polarization component $\av{S_{x}}$ is decoupled from the exciting field and the other components. The equation of motion for $\av{S_{x}}$ can be integrated immediately to give
\begin{align}
\av{S_{x}(t)} = \av{S_{x}(0)}e^{-\gamma_{x}t} .\label{H46}
\end{align}
It is a simple exponential decay with the rate $\gamma_{x}$, so in order to determine the steady-state value of $\av{S_{x}}$ we have to look at the properties of the damping rate $\gamma_{x}$. According to Eqs.~(\ref{H45}) and (\ref{H19}), the rate depends strongly on the relation between $\gamma_{1}$ and $\gamma_{2}$ and the phase $\phi$. When $\gamma_{1}>\gamma_{2}$ or $\gamma_{1}<\gamma_{2}$, at which $\gamma_{s}\neq \gamma_{n}\neq \gamma_{m}$, we see that $\av{S_{x}(t)}$ decays to zero independent of the phase. However, in the case $\gamma_{1}=\gamma_{2}$, that is when $\gamma_{s}= \gamma_{n}= \gamma_{m}$, the decay rate $\gamma_{x}$ depends on the phase. It follows that if $\phi=0$ then in the steady state $\av{S_{x}}_{s}=0$, whereas the component decays to a nonzero steady-state value $\av{S_{x}}_{s} = \av{S_{x}(0)}$ if $\phi=\pi/2$. This implies that the coherence between the ground and excited states of the quantum dot is locked at its initial value.  Hence, the two choices of phase lead to widely different behavior of the $\av{S_{x}}$ component.

The steady-state values of the remaining two components $\av{S_{y}}$ and $\av{S_{z}}$ are found to be
\begin{align}
\av{S_{y}}_{s} = \frac{(\gamma_{s}-\gamma_{n})\Omega}{\gamma_{y}\gamma_{z}+\Omega^{2}} ,\quad
\av{S_{z}}_{s} = \frac{-(\gamma_{s}-\gamma_{n})\gamma_{y}}{\gamma_{y}\gamma_{z}+\Omega^{2}} .\label{H47}
\end{align}
Provided that $\gamma_{s}\neq \gamma_{n}$, the steady-state values are different from zero. However, if $\gamma_{1}=\gamma_{2}$ we have $\gamma_{s}=\gamma_{n}$ and then $\av{S_{y}}_{s}=\av{S_{z}}_{s}=0$ regardless of the phase $\phi$. Thus, for $\gamma_{1}=\gamma_{2}$, the steady-state value of the total spin of the system depends solely on the initial value of $\av{S_{x}}$. For $\av{S_{x}(0)}\neq 0$ the polarization is locked at its initial value due to the interaction with the perfectly squeezed field.

In this way, we may modify the steady-state population distribution between dressed states of the driven quantum dot.
In order to show it, we introduce the semiclassical dressed states, which are the eigenstates of the two-level system and the classical driving field.
Since the driving laser is on resonance with the two-level transition, the dressed states are
\begin{align}
\ket{+} =\frac{1}{\sqrt{2}}(\ket{g}+\ket{e}) ,\quad
\ket{-} =\frac{1}{\sqrt{2}}(\ket{g}-\ket{e}) .\label{H48}
\end{align}
It is easily verified that the populations of the dressed states are related to the populations and coherences of the bare states through the relations
\begin{align}
\rho_{++} &= \frac{1}{2}(\rho_{11}+\rho_{22}+\rho_{12}+\rho_{21}) = \frac{1}{2}(1+2\av{S_{x}}) ,\nonumber\\
\rho_{--} &= \frac{1}{2}(\rho_{11}+\rho_{22}-\rho_{12}-\rho_{21}) = \frac{1}{2}(1-2\av{S_{x}}) .\label{H49}
\end{align}
We see that only the component $\av{S_{x}}$ contributes to the populations of the dressed states. Since in the steady-state $\av{S_{x}}_{s} =\av{S_{x}(0)}$, we see that one can polarize the dressed state populations, i.e., create an asymmetry in populations within dressed-state doublets simply by choosing an initial state at $t=0$ such that $\av{S_{x}(0)}\neq 0$.
Particularly interesting is the phenomenon of dressed state population trapping or complete polarization of the dressed state populations which happens when $\av{S_{x}} =\pm\frac{1}{2}$. From the foregoing discussion on the possible steady-state values of $\av{S_{x}}$, we see that complete polarization of the dressed state populations occurs when $\phi=\pi/2$ and initially at time $t=0$, $\av{S_{x}(0)} =\pm\frac{1}{2}$. In practice, arbitrary initial values of $\av{S_{x}}$ could be prepared using the standard technique of a $\pi/2$ pulse excitation. For example, the quantum dot could be prepared in one of the dressed states $(\av{S_{x}(0)}=\pm\frac{1}{2})$ using a $\pi/2$ pulse laser field that is $\pm\pi/2$ out of phase with the exciting field~\cite{lb86}.

Note that the polarization effect in the system considered here is obtained in the resonant case $(\Delta_{L}=0)$. This is in contrast to the polarization effect found for the steady-state of a two-level atom damped by an externally generated squeezed vacuum field, where it was shown~\cite{cr89,sd97} that the dressed-state polarization
is possible only at a non-zero laser detuning,~$\Delta_{L}\neq 0$.

\section{Fluorescence spectrum}\label{sec5}

We now consider the spectrum of the fluorescence field, which can be written as a sum of two parts
\begin{align}
S(\omega) = S_{coh}(\omega) + S_{in}(\omega) ,\label{H50}
\end{align}
where
\begin{align}
S_{coh}(\omega) = 2\pi \av{S^{+}}_{s}\av{S^{-}}_{s}\delta(\omega -\omega_{0}) \label{H51}
\end{align}
is the coherent (elastic) part of the spectrum, and
\begin{align}
S_{in} (\omega) &= 2{\rm Re}\left\{\int_0^\infty d\tau \, e^{i(\omega-\omega_{0})\tau}\right. \nonumber\\
&\left. \times \lim\limits_{t \rightarrow \infty} \av{\delta S^{+}(t)\delta S^{-}(t+\tau)}\right\} \label{H52}
\end{align}
is the incoherent (noise) part of the spectrum. Here, $\delta S^{\pm}(t)= S^{\pm}(t)-\av{S^{\pm}(t)}$ are the fluctuation parts of the dipole operators.

The two-time correlation function appearing in Eq.~(\ref{H52}) can be written as
\begin{align}
\av{\delta S^{+}(t)\delta S^{-}(t+\tau)} &= \av{\delta S^{+}(t)\delta S_{x}(t+\tau)} \nonumber\\
&-i\av{\delta S^{+}(t)\delta S_{y}(t+\tau)} ,\label{H53}
\end{align}
and the correlation functions $\av{\delta S^{+}(t)\delta S_{x}(t+\tau)}$ and $\av{\delta S^{+}(t)\delta S_{x}(t+\tau)}$ may in turn be evaluated from Eq.~(\ref{H44}) with the help of the quantum regression theorem~\cite{lax}. By Laplace transforming of the resulting equations of motion for the two-time correlation functions, we obtain
\begin{align}
\Lambda(z) &\equiv \lim\limits_{t \rightarrow \infty}{\cal L}\{\av{\delta S^{+}(t)\delta S^{-}(t+\tau)}\} = \frac{\av{\delta S^{+}\delta S_{x}}_{s}}{z+\gamma_{x}} \nonumber\\
&-i\frac{\av{\delta S^{+}\delta S_{y}}_{s}(z+\gamma_{z}) -\Omega\av{\delta S^{+}\delta S_{z}}_{s}}{z^{2}+(\gamma_{y}+\gamma_{z})z+\gamma_{y}\gamma_{z} +\Omega^{2}} ,\label{H54}
\end{align}
where $z$ is the Laplace transform parameter and the steady-state averages of the various operator products arising are
\begin{align}
\av{\delta S^{+}\delta S_{x}}_{s} &= \frac{1}{2}\!\left(\!\av{S_{z}}_{s}+\frac{1}{2}\right) -\av{S_{x}}_{s}\left(\av{S_{x}}_{s}+i\av{S_{y}}_{s}\right) ,\nonumber\\
\av{\delta S^{+}\delta S_{y}}_{s} &= \frac{i}{2}\!\left(\!\av{S_{z}}_{s}+\frac{1}{2}\right) -\av{S_{y}}_{s}\left(\av{S_{x}}_{s}+i\av{S_{y}}_{s}\right) ,\nonumber\\
\av{\delta S^{+}\delta S_{z}}_{s} &= -\frac{1}{2}(\av{S_{x}}_{s}+i\av{S_{y}}_{s})\left(1+2\av{S_{z}}_{s}\right).\label{H55}
\end{align}

To illustrate the analytic structure of the spectrum in as simple form as possible, we focus on the case $\gamma_{1}=\gamma_{2}$ and the strong-field limit $\Omega\gg \gamma_{1},\gamma_{2}$.
The cases, $\gamma_{1}>\gamma_{2}$ and $\gamma_{2}>\gamma_{1}$ lead to the spectrum which is essentially the same as that of the fluorescence field emitted by a two-level system whose relaxation is determined through coupling to a squeezed vacuum field produced by an external source~\cite{cl87,pa90,df99,te16}.
In the strong-field limit, we readily find that the two roots of the quadratic equation in the denominator of Eq.~(\ref{H54}) are approximately given by
\begin{align}
z_{1,2} = -\frac{1}{2}(\gamma_{y}+\gamma_{z})\pm i\Omega ,\label{H56}
\end{align}
and then the spectral components take the form
\begin{align}
S_{coh}(\omega) = 2\pi \av{S_{x}}^{2}_{s}\delta(\omega -\omega_{0}) ,\label{H57}
\end{align}
and
\begin{align}
S_{in}(\omega) &= 2{\rm Re}\{\Lambda(z)\}_{z=-i(\omega -\omega_{0})} \nonumber\\
&=\frac{1}{2}\!\left(1-4\av{S_{x}}^{2}_{s}\right)\frac{\gamma_{x}}{\gamma_{x}^{2}+(\omega -\omega_{0})^{2}} \nonumber\\
&+\!\frac{\frac{1}{8}\left(1+2\av{S_{x}}_{s}\right)(\gamma_{y}+\gamma_{z})+\frac{\gamma_z-\gamma_y}{8\Omega}(\omega-\omega_0-\Omega)}{\frac{1}{4}(\gamma_{y}\!+\!\gamma_{z})^{2}+(\omega -\omega_{0}-\Omega)^{2}} \nonumber\\
&+\!\frac{\frac{1}{8}\left(1-2\av{S_{x}}_{s}\right)(\gamma_{y}+\gamma_{z})+\frac{\gamma_z-\gamma_y}{8\Omega}(\omega-\omega_0+\Omega)}{\frac{1}{4}(\gamma_{y}\!+\!\gamma_{z})^{2}+(\omega -\omega_{0}+\Omega)^{2}} .\label{H58}
\end{align}
where we have used the fact that $\av{S_{y}}_{s}=\av{S_{z}}_{s}=0$.

One can see from Eqs.~(\ref{H57}) and (\ref{H58}) that the amplitudes of the spectral components are solely dependent on the polarization (coherence) component $\av{S_{x}}_{s}$.
We first note that the coherent part of the spectrum is present only if $\av{S_{x}}_{s}\neq 0$. Otherwise when $\av{S_{x}}_{s}=0$ the spectrum consists only of the incoherent part, which is always present. In general, the incoherent part of the spectrum is composed of three Lorentzian peaks of the widths and magnitudes varying with the phase $\phi$.
The most interesting feature of the incoherent spectrum is its asymmetry related to $\av{S_{x}}_{s}\neq 0$, because this feature is not encountered at all under the damping of the quantum dot by an externally produced squeezed vacuum field. If we consider the variation of the spectrum with the phase, we find for~$\phi=0$,
\begin{align}
\gamma_{x}= \gamma_{z} = 4(2\bar{n}+1)\gamma_{0} ,\quad \gamma_{y}=0 ,\quad \av{S_{x}}_{s}=0 ,\label{H59}
\end{align}
while for $\phi=\pi/2$,
\begin{align}
\gamma_{x}= 0 ,\ \gamma_{y} = \gamma_{z} = 4(2\bar{n}+1)\gamma_{0} ,\ \av{S_{x}}_{s}= \av{S_{x}(0)} .\label{H60}
\end{align}
For $\phi=0$, the coherent part of the spectrum is suppressed, whereas the incoherent part is composed of three peaks of equal amplitudes. The width of the central peak is $4(2\bar{n}+1)\gamma_{0}$ and it is twice as wide as the width of the Rabi sidebands. Thus, the spectrum is symmetric about the laser frequency and entirely composed of the incoherent part.

The spectrum changes dramatically when the phase is varied to $\phi=\pi/2$. The coherent part appears and the central peak of the incoherent part becomes a $\delta$-type peak. The widths of the Rabi sidebands are twice as wide as for $\phi=0$. Thus, for $\phi=\pi/2$, the central peak contributes to a coherent (elastic) part of the spectrum leading to an enhancement of the coherent scattering. The incoherent part is then effectively composed of two peaks located at the Rabi sidebands. Depending on $\av{S_{x}(0)}$ the number of peaks in the incoherent part can vary from three to a single side peak located at $\omega-\omega_{0}=\pm\Omega$. The disappearance of two peaks is a consequence of the complete polarization of the dressed state population. For example, when $\av{S_{x}(0)}=\frac{1}{2}$, the population is entirely in the upper dressed state $\ket{+}$. Consequently, the central and the lower frequency Rabi sideband peaks disappear. On the other hand, when $\av{S_{x}(0)}=-\frac{1}{2}$, the population is entirely in the lower dressed state $\ket{-}$ resulting in the absence of the central peak and the higher frequency Rabi sideband. The disappearance of the central peak is accompanied by an increase of the amplitude of the coherent part of the spectrum. In other words, the energy contained in the central peak is coherently scattered by the quantum dot. The disappearance of one of the Rabi sidebands is accompanied by an increase of the amplitude of the opposite Rabi sideband, which after the complete transfer of the population is twice as high as for $\av{S_{x}(0)}=0$.
\begin{figure}[h]
\centering\includegraphics[width=\columnwidth,keepaspectratio,clip]{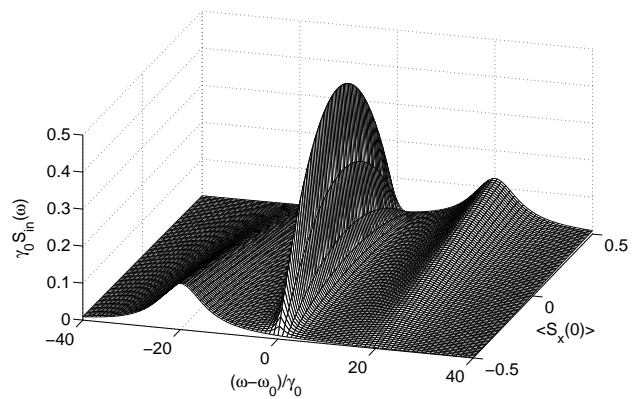}
\caption{The incoherent part of the fluorescence spectrum as a function of $\av{S_{x}(0)}$ for $\phi=\pi/2$, $\Omega=20\gamma_{0}$ and $\bar{n}=0.5$. In order to make the central component visible in the plot, the delta function has been replaced by a Lorentzian of a width $\gamma_{0}$. }
\label{bfig5}
\end{figure}

The features described above are easily seen in Fig.~\ref{bfig5} which shows the incoherent fluorescence spectrum plotted as a function of $\av{S_{x}(0)}$ for fixed $\Omega$ and $\bar{n}$. We see the disappearance of the central peak and one of the Rabi sidebands when $\av{S_{x}(0)}=\pm\frac{1}{2}$, and simultaneously the increase in the height of the opposite Rabi sideband. We again point out that the asymmetric spectrum and its variation with the initial coherence are obtained under strictly resonant excitation. The variation of the fluorescence spectrum with the initial coherence shows clearly that the phase relationships between the irradiating field and the initial dipole moment of the radiating quantum dot are important even in the steady-state fluorescence.

\section{Summary}\label{sec6}

We have shown that the combined effect of a phonon bath and a bichromatic field can result in an effective squeezed-vacuum-type reservoir to a two-level quantum dot.
It has been found that depending on the ratio of the Rabi frequencies of the components of the bichromatic field, one can engineer a squeezed reservoir of ordinary or inverted harmonic oscillators. We have shown that in the case of the inverted harmonic oscillators the steady-state population of the bare states of the quantum dot can be highly inverted. We have examined the conditions for quantum two-photon correlations characteristic of a squeezed field and have distinguished between the quantum correlations and a form of classical two-photon correlations which may exist at high temperatures of the phonon bath.

When in addition to the squeezed reservoir, the quantum dot interacts with a strong laser field, the dynamics and the stationary state could depend critically on whether or not a coherence between the ground and excited states is initially present. With the coherence present, the steady-state population distribution between the dressed states of the driven quantum dot could be completely polarized (trapped) in one of the dressed states. We have calculated the steady-state spectrum of the fluorescence field and have found that the structure spectrum depends on the initial value of the coherence. In particular, with a nonzero initial coherence the spectrum is asymmetric even if the quantum dot is exposed to a resonant laser field. We have found that the asymmetric features are the same as those exhibited by the spectrum of a two-level system excited off resonance and damped by a squeezed vacuum field produced by an external squeezing source.
The appearance of the asymmetric features have been interpreted as a direct consequence of locking the coherence at its initial value, resulting from the coupling of the quantum dot to the perfectly squeezed field.  In the absence of the initial coherence, the spectrum was found to be composed of only the incoherent part displaying the symmetric triplet spectrum. With the coherence present, both coherent and incoherent parts are present and the number of peaks in the incoherent part becomes strongly dependent on phase. By varying the phase, the central peak can become a $\delta$-type peak and one of the Rabi sidebands could be suppressed.

\section*{Acknowledgments}
This work was supported by the National Natural Science Foundation of China (Grants No. 61275123, No. 11474119), and the National Basic Research Program of China (Grant No. 2012CB921602).


\begin{thebibliography}{90}

\bibitem{rm08} E. Rozbicki and P. Machnikowski, Phys. Rev. Lett. {\bf 100}, 027401 (2008).
\bibitem{mb12}  A. Majumdar, M. Bajcsy, A. Rundquist, E. Kim, and J. Vu\v{c}kovi\'{c}, Phys. Rev. B {\bf 85}, 195301 (2012).
\bibitem{rgg10} A. J. Ramsay, A. V. Gopal, E. M. Gauger, A. Nazir, B. W. Lovett, A. M. Fox, and M. S. Skolnick, Phys. Rev. Lett. {\bf 104}, 017402 (2010).
\bibitem{mt13}  L. Monniello, C. Tonin, R. Hostein, A. Lemaitre, A. Martinez, V. Voliotis, and R. Grousson, Phys. Rev. Lett. {\bf 111}, 026403 (2013).
\bibitem{fw03} J. F\"{o}rstner, C. Weber, J. Danckwerts, and A. Knorr, Phys. Rev. Lett. {\bf 91}, 127401 (2003).
\bibitem{ah04} R. Alicki, M. Horodecki, P. Horodecki, R. Horodecki, L. Jacak, and P. Machnikowski, Phys. Rev. A {\bf 70}, 010501(R) (2004).
\bibitem{mn10} D. P. S. McCutcheon and A. Nazir, New J. Phys. {\bf 12}, 113042 (2010).
\bibitem{vc11} A. Vagov, M. D. Croitoru, M. Gl\"{a}ssl, V.M. Axt, and T. Kuhn, Phys. Rev. B {\bf 83}, 094303 (2011).
\bibitem{md11} D. P. S. McCutcheon, N. S. Dattani, E. M. Gauger, B. W. Lovett, and A. Nazir, Phys. Rev. B {\bf 84}, 081305(R) (2011).
\bibitem{mp12} A. Moelbjerg, P. Kaer, M. Lorke, and J. M{\o}rk, Phys. Rev. Lett. {\bf 108}, 017401 (2012).
\bibitem{mn13} D. P. S. McCutcheon and A. Nazir, Phys. Rev. Lett. {\bf 110}, 217401 (2013).
\bibitem{wu12} S. Weiler, A. Ulhaq, S. M. Ulrich, D. Richter, M. Jetter, P. Michler, C. Roy, and S. Hughes, Phys. Rev. B {\bf 86}, 241304 (2012).
\bibitem{zh13} J.-P. Zhu, H. Huang, and G.-X. Li, Phys. Rev. A {\bf 88}, 023835 (2013).
\bibitem{zh14} J.-P. Zhu, H. Huang, and G.-X. Li, J. Appl. Phys. {\bf 115}, 033102 (2014).
\bibitem{yc11} W.-X. Yang, A.-X. Chen, R.-K. Lee, and Y. Wu, Phys. Rev. A {\bf 84}, 013835 (2011).

\bibitem{ka06} A. Kr\"{u}gel, V. M. Axt, and T. Kuhn, Phys. Rev. B {\bf 73}, 035302 (2006).
\bibitem{vc07} A. Vagov, M. D. Croitoru, V. M. Axt, T. Kuhn, and F. M. Peeters, Phys. Rev. Lett. {\bf 98}, 227403 (2007).
\bibitem{fm09} E. B. Flagg, A. Muller, J. W. Robertson, S. Founta, D. G. Deppe, M. Xiao, W. Ma, G. J. Salamo, and C. K. Shih, Nature Physics {\bf 5}, 203 (2009).
\bibitem{vz09} A. N. Vamivakas, Y. Zhao, C.-Y. Lu, and M. Atat\"{u}re, Nature Physics {\bf 5}, 198 (2009).
\bibitem{rg10} A. J. Ramsay, T. M. Godden, S. J. Boyle, E. M. Gauger, A. Nazir, B. W. Lovett, A. M. Fox, and M. S. Skolnick, Phys. Rev. Lett. {\bf 105}, 177402 (2010).
\bibitem{ua11} S. M. Ulrich, S. Ates, S. Reitzenstein, A. L\"{o}ffler, A. Forchel, and P. Michler, Phys. Rev. Lett. {\bf 106}, 247402 (2011).
\bibitem{rh11} C. Roy and S. Hughes, Phys. Rev. Lett. {\bf 106}, 247403 (2011).
\bibitem{rh12} C. Roy and S. Hughes, Phys. Rev. X {\bf 1}, 021009 (2011).
\bibitem{hh15} Y. He, Y. M. He, J. Liu, Y.-J. Wei, H. Y. Ram\'{\i}rez, M. Atat\"{u}re, C. Schneider, M. Kamp, S. H\"{o}fling, C.-Y. Lu, and J.-W. Pan, Phys. Rev. Lett. {\bf 114}, 097402 (2015).
\bibitem{uw13} A. Ulhaq, S. Weiler, C. Roy, S. M. Ulrich, M. Jetter, S. Hughes, and P. Michler, Opt. Express {\bf 21}, 4382 (2013).
\bibitem{wh13} Y.-J. Wei, Y. He, Y.-M. He, C.-Y. Lu, J.-W. Pan, C. Schneider, M. Kamp, S. H\"{o}fling, D. P. S. McCutcheon, and A. Nazir, Phys. Rev. Lett. {\bf 113}, 097401 (2014).


\bibitem{kamp} S. Unsleber, S. Maier, D. P. S. McCutcheon, Y.-M. He, M. Dambach, M. Gschrey, N. Gregersen, J. M{\o}rk, S. Reitzenstein, S. H\"{o}fling, C. Schneider, and M. Kamp, Optica {\bf 2}, 1072 (2015).

\bibitem{hc13} S. Hughes and H. J. Carmichael, New J. Phys. {\bf 15}, 053039 (2013).
\bibitem{macovei} S. Das and M. A. Macovei, Phys. Rev. B {\bf 88}, 125306 (2013).

\bibitem{huang} H. Huang, G.-X. Li, W.-J. Gu, and Z. Ficek, Phys. Rev. A  {\bf 90}, 023815 (2014).
\bibitem{simon} C.-M. Simon, T. Belhadj, B. Chatel, T. Amand, P. Renucci, A. Lemaitre, O. Krebs, P. A. Dalgarno, R. J. Warburton, X. Marie, and B. Urbaszek, Phys. Rev. Lett. {\bf 106}, 166801 (2011).
\bibitem{wu} Y. Wu, I. M. Piper, M. Ediger, P. Brereton, E. R. Schmidgall, P. R. Eastham, M. Hugues, M. Hopkinson, and R. T. Phillips, Phys. Rev. Lett. {\bf 106}, 067401 (2011).

\bibitem{singlephoton1} A. Imamo\={g}lu and Y. Yamamoto, Phys. Rev. Lett. {\bf 72}, 210 (1994).

\bibitem{singlephoton2} T. Grange, G. Hornecker, D. Hunger, J.-P. Poizat, J.-M. G\'{e}rard, P. Senellart, and A. Auff\`{e}ves, Phys. Rev. Lett. {\bf 114}, 193601 (2015).

\bibitem{sci} J. Tang, W. Geng, and X. Xu, Scientific Rep. {\bf 5}, 9252 (2015).

\bibitem{ding} X. Ding, Y. He, Z.-C. Duan, N. Gregersen, M.-C. Chen, S. Unsleber, S. Maier, C. Schneider, M. Kamp, S. H\"{o}fling, C.-Y. Lu, and J.-W. Pan, Phys. Rev. Lett. {\bf 116}, 020401 (2016).

\bibitem{na08} A. Nazir, Phys. Rev. B {\bf 78}, 153309 (2008).


\bibitem{dressed1} A. Majumdar, M. Bajcsy, and J. Vu\v{c}kovi\'{c}, Phys. Rev. A {\bf 85}, 041801(R) (2012).

\bibitem{gao} B. Gao, J.-P. Zhu, and G.-X. Li, J. Appl. Phys. {\bf 119}, 103104 (2016).


\bibitem{ga86} C. W. Gardiner, Phys. Rev. Lett. {\bf 56}, 1917 (1986).
\bibitem{cl87} H. J. Carmichael, A. S. Lane, and D. F. Walls, Phys. Rev. Lett. {\bf 58}, 2539 (1987); J. Mod. Opt. {\bf 34}, 821 (1987).
\bibitem{pa90} A. S. Parkins, Phys. Rev. A  {\bf 42}, 6873 (1990).
\bibitem{df99} B. J. Dalton, Z. Ficek, and S. Swain, J. Mod. Opt. {\bf 46}, 379 (1999).
\bibitem{df04}  P. D. Drummond and Z. Ficek (editors), {\it Quantum Squeezing}, (Springer, New York, 2004).

\bibitem{gp95} N. Ph. Georgiades, E. S. Polzik, K. Edamatsu, H. J. Kimble, and A.S. Parkins, Phys. Rev. Lett. {\bf 75}, 3426 (1995).
\bibitem{gp97} N. Ph. Georgiades, E. S. Polzik, and H. J. Kimble, Phys. Rev. A {\bf 55}, R1605 (1997).
\bibitem{wk86} L. A. Wu, H. J. Kimble, J. L. Hall, and H. Wu, Phys. Rev. Lett. {\bf 57}, 2520 (1986).
\bibitem{wx87} L. A. Wu, M. Xiao, and H. J. Kimble, J. Opt. Soc. Am. B {\bf 4}, 1465 (1987).
\bibitem{mw13} K. W. Murch, S. J. Weber, K. M. Beck, E. Ginossar, and I. Siddiqi, Nature (London) {\bf 499}, 62 (2013).
\bibitem{te16} D. M. Toyli, A. W. Eddins, S. Boutin, S. Puri, D. Hover, V. Bolkhovsky, W. D. Oliver, A. Blais, and I. Siddiqi, Phys. Rev X {\bf 6}, 031004 (2016).
\bibitem{tg98} Q. A. Turchette, N. Ph. Georgiades, C. J. Hood, H. J. Kimble, and A. S. Parkins, Phys. Rev. A {\bf 58}, 4056 (1998).

\bibitem{pc96} J. F. Poyatos, J. I. Cirac, and P. Zoller, Phys. Rev. Lett. {\bf 77}, 4728 (1996).
\bibitem{cp03} S. Clark, A. Peng, M. Gu, and S. Parkins, Phys. Rev. Lett. {\bf 91}, 177901 (2003).
\bibitem{lc98} N. L\"{u}tkenhaus, J. I. Cirac, and P. Zoller, Phys. Rev. A {\bf 57}, 548 (1998).

\bibitem{hp11} Q. Y. He, S. G. Peng, P. D. Drummond, and M. D. Reid, Phys. Rev. A {\bf 84}, 022107 (2011).
\bibitem{hv12} Q. Y. He, T. G. Vaughan, P. D. Drummond, and M. D. Reid, New J. Phys. {\bf 14}, 093012 (2012).
\bibitem{dg15} B. J. Dalton, J. Goold, B. M. Garraway, and M. D. Reid, arXiv:1506.06892.

\bibitem{wm85} D. F. Walls and G. J. Milburn, Phys. Rev. A {\bf 31}, 2403 (1985).

\bibitem{lb86} N. Lu, P. R. Berman, A. G. Yodh, Y. S. Bai, and T. W. Mossberg, Phys. Rev. A {\bf 33}, 3956 (1986).
\bibitem{cr89} J. M. Courty and S. Reynaud, Europhys. Lett. {\bf 10}, 237 (1989).
\bibitem{sd97} S. Swain and B. J. Dalton, Optics Commun. {\bf 147}, 187 (1998).

\bibitem{lax} M. Lax, Phys. Rev. {\bf 157}, 213 (1967).

\end{thebibliography}
\end{document}